\documentclass[12pt]{article}

\usepackage{amsmath}
\usepackage{amssymb}
\usepackage[latin2]{inputenc}
\usepackage{graphicx}

\newcommand{\<}{\langle}
\renewcommand{\>}{\rangle}
\def\Prob{{\rm Prob\,}}
\def\Det{{\rm Det\,}}

\def\bbbc{{\mathbb C}}

\def\Tr{\mbox{Tr}\,}

\def\iiint{\int\!\!\int\!\!\int}

\title{{\bf Comparison of some methods of quantum state 
estimation\footnote{To be published in the proceedings of the
"26th Conference: QP and IDA" - Levico (Trento), February, 2005.}
}}
\author{Th. Baier$^1$\footnote{Supported by the EU
Research Training Network Quantum Probability with Applications to
Physics, Information Theory and Biology.}, $\,\,$
K.M. Hangos$^2$\footnote{Supported by the Hungarian grant OTKA T042710.}, 
$\,\,$
A. Magyar$^2$ and
D. Petz$^1$\footnote{Supported by the Hungarian grant OTKA T032662.} 
\\ \phantom{mm} \\
{\small $^1$Department for Mathematical Analysis, Budapest University 
of Technology and Economics} \\
{\small H-1111 Budapest, M\H uegyetem rkp. 2-8, Hungary}\\ \phantom{mm} \\
{\small $^2$Process Control Research Group, Computer and Automation Research Institute} \\
{\small H-1518 Budapest, POBox 63, Hungary}
}

\bigskip

\date{}

\begin{document}

\maketitle

\begin{abstract}
In the paper the Bayesian and the least squares methods of quantum
state tomography are compared for a single qubit. The quality of
the estimates are compared by computer simulation when the true
state is either mixed or pure. The fidelity and the Hilbert-Schmidt distance
are used to quantify the error.

It was found that in the regime of low measurement number the Bayesian method
outperforms the least squares estimation. Both methods are quite sensitive to 
the degree of mixedness of the state to be estimated, that is, their performance
can be quite bad near pure states.

\end{abstract}

\section{Introduction}
The aim of quantum state estimation is to decide the actual state of a
quantum system by measurements. Since the outcome of a measurement is
stochastic, several measurements are to be done and statistical
arguments lead to the reconstruction of the state. Due to some
similarities with X-ray tomography, the state reconstruction is
often called {\it quantum tomography}\cite{GillGuta:2005}.
More precisely, in physics-related
books, journals and papers, tomography refers to both the state and
parameter estimation of quantum dynamical systems, the term
{\it state  tomography} is used for the first, and {\it process tomography}
is applied for the second
 case\cite{NielsenChuang:2000,ArPaSa:2003,DarMacPar:2000}.
The engineering literature contains
also papers related to state and parameter estimation of quantum systems
but they term it \emph{identification} for the case of parameter
estimation\cite{KoWaRa:2005,AlbDAl:2005}  and
\emph{state filtering} for the case of state estimation\cite{KoWaRa:2004}.

In this paper the estimation of the state of a qubit is discussed.
This is the simplest possible case of quantum state estimation where no
dynamics is assumed and the measurements are performed on identical copies of
the qubit. Therefore, the state estimation problem reduces to a static
parameter estimation problem, where the parameters to be estimated are
the parameters of the density matrix of the qubit.

The methods of \emph{classical statistical estimation} are used to
 develop
 state estimation of quantum systems in the first group of
 papers\cite{Hel:1976,ArPaSa:2003,ReEnKA:2004}.
This approach suffers from the fact that the state estimation
 is usually based on a few types of measurement (observables)
 that are incompatible, thus there is no joint probability
 density function of the measurement results in the
 classical sense\cite{Hradil:1999}.

The most common way of statistical state estimation is
 the maximum-likelihood (ML) method that leads to a convex optimization
 problem in the qubit case (see below). The \emph{convex optimization 
methods} are used in other approaches as well, see 
\cite{KoWaRa:2004,KoWaRa:2005}.
Here one can respect the constraints imposed on the components of
 the state but there is no information on the probability distribution of
 the estimate.

The efficiency of the ML estimate, its asymptotic properties and the
 Cram\'er-Rao bound can be used to derive consequences on the asymptotic
  distribution of an estimate and on its variance.
This approach has been used for optimal experiment design
 in\cite{KoWaRa:2004}. A lower bound on the estimation error for qubit
state estimation is derived in\cite{HaMa:2004}.

It is natural to require that any state estimation scheme should be
 unbiased and should converge in some stochastic sense to the true
 value if the number of samples (measurements done) tends to infinity.
The basis of the comparison is then a suitably chosen \emph{measure of fit}
 (for example averaged fidelities with respect to the true density matrix, or
 the variance of the estimate).
The fidelity and the Bures-metric defined therefrom was used to
 derive optimal estimators of qubit state in\cite{BaBa:2003}.
Fidelity has also been used to evaluate the performance of an estimation
 scheme\cite{BaBa:2005} for the so called "purity" of a qubit (i.e.\
the length of its
 Bloch vector) in the context of Bayesian state estimation.

Large deviations can also be used to analyze the performance of state
estimation schemes\cite{Keyl:2004}, when the qubit is in a mixed state.
An optimal estimation scheme is also proposed based on covariant observables.

The aim of this paper is to investigate the properties of two state estimation
 methods, the Bayesian state estimation as a statistical method and the least
 squares (LS) method as an optimization-based method by using simulation
experiments.
The simplest possible quantum system, a single qubit, a quantum two level
system, is applied, where we could compute some of the
estimates analytically.

\section{Preliminaries about two level systems}

The general state of a two level quantum system is described
by a density operator $\rho$, which is a positive operator on
the Hilbert space $\bbbc^2$,  normalized to $\Tr \rho = 1$.
On the one hand, $\rho$ is represented in the form of a $2
\times 2$ matrix, and on the other hand by the so-called
{\it Bloch vector}
$
    {s}=[s_1,s_2,s_3]^T
$.
With use of the Pauli matrices
$$
\sigma_1 =\left[ \begin{array}{cc}
                          0 & 1 \\
                          1 & 0
                     \end{array}
\right], \quad
    \sigma_2 =\left[\begin{array}{cc}
                          0 & -i \\
                          i & \phantom{-}0
                     \end{array}
    \right],\quad
    \sigma_3 =\left[\begin{array}{cc}
                      1 & \phantom{-}0 \\
              0 & -1
                    \end{array}
    \right],
$$
the correspondence between the density operator $\rho$ and the Bloch vector
${s}$ is given by the expansion
$$
\rho = \frac12 (I + s_1 \sigma_1 + s_2 \sigma_2 + s_3 \sigma_3),
$$
where the constraint
\begin{equation}\label{constraint}
    \|{s}\|= \sqrt{s_1^2 + s_2^2 + s_3^2} \le 1
\end{equation}
is satisfied. The correspondence between $\rho$ and ${s}$ is affine.
Thus the state space of a spin system is represented by the three
dimensional unit ball, called the {\it Bloch ball}.

Observables, i.e. physical quantities to be measured, are
represented by self-adjoint operators acting on the underlying
Hilbert space\cite{JvN:1957}. A self-adjoint operator $A$ has a spectral
decomposition
$
    A = \sum_{i=1}^n \lambda_i P_i
$.
The different eigenvalues $\lambda_i$ of the operator $A$ correspond to the possible
outcomes of the measurement of the associated observable and the $i$th outcome
occurs  with probability
$
    \Prob(\lambda_i)= \Tr \rho P_i
$,
where $P_i$ is the projection onto the subspace of the corresponding
eigenvectors. Consequently, the expectation value of the measurement is
$$
\< A \>_\rho : = \sum_i \lambda_i \Prob(\lambda_i)= \Tr \rho A.
$$

\section{Measurements on  qubits}

For the state estimation, we will consider $3n$ identical
copies of qubits in the state $\rho$. On each copy in this passel, we perform
a measurement of one of the Pauli spin matrices $\{\sigma_1,\sigma_2, \sigma_3\}$,
each of them $n$ times. The possible outcomes for each of this single
measurements, i.e. the eigenvalues of the $\sigma_i$, are $\pm 1$ and the
corresponding spectral projections are given by
\begin{equation} \label{projections}
    P^{\pm}_i= \frac12 (I \pm \sigma_i).
\end{equation}
For the sake of definiteness, we assume that first $\sigma_1$ is measured $n$ times,
then $\sigma_2$ and then $\sigma_3$.
The {\it data set} of the outcomes of this measurement scheme consists
of three strings of length $n$ with entries $\pm 1$:
\begin{equation}\label{E:data}
        D_i^n=\{D_i^n(j):j=1, \cdots,n\} \qquad (i=1,2,3).
\end{equation}
The predicted probabilities of the outcomes depend on
the true state $\rho$ of the system and they are given by
\begin{equation}\label{prob_outcome}
    \Prob(D_i^n(j)=1) = \Tr (\rho P_i^+) = \frac12 (1 + \< \sigma_i\>_\rho)
    = \frac12 (1+s_i).
\end{equation}

\section{Quality of the estimates}

As a measure of distance between two states of a system, i.e.\
 between two density operators $\rho$ and $\omega$, the
{\it fidelity}
\begin{equation}\label{E:fidelity}
    F(\rho,\omega) = \Tr \sqrt{\rho^{\frac12} \omega \rho^{\frac12}}
\end{equation}
can be considered\cite{Uhlmann:1976,NielsenChuang:2000}. It fulfills
the properties
\begin{eqnarray*}
        F(\rho,\omega) = F(\omega, \rho), \quad  0\le F(\rho,\omega)\le 1 \\
    F(\rho,\omega) = 1 \iff \rho = \omega,\quad
    F(\rho,\omega) = 0 \iff\omega \perp \rho\,.
\end{eqnarray*}
For spin 1/2 systems the fidelity can be calculated from the eigenvalues
$\lambda_{1}$ and $\lambda_{2}$ of the operator $A = \rho^{\frac12} \omega
\rho^{\frac12}$ as
\begin{equation*}
        F(\rho,\omega) = \sqrt{\lambda_1} + \sqrt{\lambda_2}.
\end{equation*}
These eigenvalues can be computed from $\Tr A$ and $\Det(A)$ as
\begin{equation*}
        \lambda_{1,2} = \frac12 \Tr A \pm \sqrt{ \frac14 \Tr A - \Det A}.
\end{equation*}
If we express $\Tr A$ and $\Det A$ in terms of the Bloch vectors ${s}$
(resp. ${r}$) of $\rho$ (resp. $\omega$), the fidelity can be written as
\begin{equation}
     F(\rho,\omega) = \frac12 \left(\sqrt{1 + {r} \cdot {s}
          + T}-
      \sqrt{1 + {r}\cdot {s} - T}\right),
\end{equation}
where
$$
T=\sqrt{\| r + s \|^2 + (r \cdot s)^2 -
          \|r\|^2 \|s\|^2 }.
$$

The quality of the estimation scheme for a true state $\rho$ can be
quantified by the average fidelity between the true state and the
estimates $\omega_i$ ($1 \le i \le m$):
$$
    \Phi (\rho,m): = \frac1{m} \sum_{i=1}^m F(\rho,\omega_i ).
$$
if $m$ estimates are available.

Alternatively, the {\it Hilbert-Schmidt distance}
\begin{equation}\label{E:HS}
d(\rho,\omega):=\sqrt{\Tr (\rho - \omega)^2}
\end{equation}
can be used as a measure. In terms of the Bloch vectors, this reduces to
$
\sqrt{\sum_i (s_i - r_i)^2}
$.
The average Hilbert-Schmidt distance is given by
$$
\chi (\rho,m): = \frac{1}{m}\sum_{i=1}^m  d(\rho, \omega_i).
$$

Remember that for an efficient estimation scheme $\chi (\rho,m)$ must
be small, while $\Phi (\rho,m)$ should be close to 1.

\section{Bayesian state estimation}

First we give a brief summary of the Bayesian state estimation. In
the Bayesian parameter estimation, the parameters $\theta$ to be
estimated are considered as random variables. The probability
$
      P(\theta~|~D^n)
$
of a specific value of the parameters conditioned on the measured
data $D^n$ is evaluated. Afterwards, the mean value of this
distribution can be used as the estimate.

If the measured data is a sequence of outcomes, as in our case, it
can be split into the latest outcome $D^n(n)$ of $D^n$ and $D^{n-1}$,
the preceding. Then the conditional distribution of the
parameter becomes
$$
     P(\theta~|~D^n(n),D^{n-1})
$$
and the Bayes formula
$$
    P(a|b, c) = \frac{P(b|a, c)P(a|c)}{ \int P(b|\nu, c)P(\nu|c)\, d\nu}
$$
can be applied resulting in the following recursive formula
for $P(\theta~|~D^n)$
\begin{equation} \label{Bayes_formula}
P(\theta~|~D^n) =
\frac{P(D^n(n)~|~D^{n-1},~\theta)P(\theta~|~D^{n-1})}
 {\int P(D^n(n)~|~D^{n-1},~\nu) P(\nu~|~D^{n-1}) d \nu }.
\end{equation}

In our state estimation, we have three data sets $D_i^n$,
$i=1,2,3$, corresponding to the three directions, see
(\ref{E:data}). The estimation is performed for the three
directions independently (and afterwards a conditioning has to be
made).

The probabilities $P(D_i^n(n)~|~D_i^{n-1},~\theta)$ have the form
$$
        P(D_i^n(n)~|~D_i^{n-1},~s_i)=P(\pm 1 ~|~s_i)=
\frac12 \Tr \rho (1 \pm \sigma_i)=\frac12 (1 \pm s_i).
$$
If we denote by $\ell(i)$ the number of $+1$'s in the data string $D_i^n$, then
(\ref{Bayes_formula}) becomes
\begin{equation} \label{likelihood_s1}
P(s_i|D_i^n)(t)=\frac{(\frac{1}{2}(1+t))^{\ell(i)}(\frac{1}{2}
(1-t))^{n-\ell(i)}P_i^0(t)}
{\int(\frac{1}{2}(1+\nu))^{\ell(i)}(\frac{1}{2}(1-\nu))^{n-\ell(i)}P_i^0(\nu)\,d\nu}
\end{equation}
where $P_i^0(\nu)$ is an assumed {\it prior distribution}, from which the recursive
estimation is started. For the sake of simplicity we assume that $P_i^0(\nu)$
has similar form with parameters $\kappa$ and $\lambda$ in place of $n$ and $\ell$,
respectively. (These parameters might depend on $i$, but we neglect this
possibility.)

After a parameter transformation we have a {\it beta distribution},
\begin{equation} \label{likelihood_s12}
P(s_i|D_i^n)(u)=C \left(\frac{1+u}{2}\right)^{\ell(i)+\lambda}
\left(\frac{1-u}{2}\right)^{n+\kappa-\ell(i)-\lambda}
\end{equation}
where $C$ is the normalization constant and $u \in [0,1]$.
It is well-known that the mean
value of this distribution is
\begin{equation} \label{est_1}
 m_i = \frac{\ell(i)+1+\lambda}{n+\kappa+2}
\end{equation}
and the variance is
\begin{equation} \label{E:variance}
\frac{(\ell(i)+1+\lambda)(n-\ell(i)+1+\kappa-\lambda)}{(n+\kappa+2)^2(n+\kappa+3)}.
\end{equation}

The above statistics (\ref{est_1}) can be used to construct an unbiased estimate
for $s_i$ in the form
\begin{equation} \label{est_s}
 \hat{s}_i = 2 \, \frac{\ell(i)+1+\lambda}{n+\kappa+2} -1
\end{equation}
after the re-transformation of the variables.

Since the components of the Bloch vector are estimated independently, the
constraint (\ref{constraint}) has not been taken into account yet. 
Thus, a further
step of conditioning is necessary. We simply condition $(\hat{s}_1,
 \hat{s}_2,  \hat{s}_3)$ to (\ref{constraint}):
\begin{equation} \label{mean_reg}
         \tilde{m_i} =\frac{
\iiint u_i f(u_1)f(u_2)f(u_3)\,du_1\,du_2\,du_3}{
\iiint f(u_1)f(u_2)f(u_3)\,du_1\,du_2\,du_3},
\end{equation}
where both integrals are over the domain $\{(u_1,u_2,u_3):
u_1^2+u_2^2+u_3^2 \le 1\}$ and
$$
f(u_i):=P(s_i|D_i^n)(u_i)\, .
$$
Then the conditioned estimate of $s_i$ will be
$$
2 \tilde{m_i}-1\,.
$$

The justification of the proposed conditioning procedure is the subject
of another publication.

\section{Least squares state estimation}

We have the {\it data set} (\ref{E:data}) to start with. If $\pi_i(\pm)$
is the relative frequency of $\pm 1$ in the string $D_i^n$, then the difference
$$
\pi_i:=\pi_i(+)-\pi_i(-)
$$
is an estimate of the $i$th spin component $s_i$ ($i=1,2,3$).
As a measure of \emph{unfit} (estimation error) we use the Hilbert-Schmidt norm of
the difference between the empirical and the predicted data according
to the least squares (LS) principle. (Note that in this case the Hilbert-Schmidt norm
is simply the Euclidean distance in the 3-space.)
Then the following loss function is defined:
\begin{equation}  \label{eq_loss}
  L(\omega) \equiv d^2({s}, {\pi}) = \sum_{j=1}^3 \left(s_{j}- \pi_j\right)^2
                             = \|{s}\|^2 + \|{\pi}\|^2 - 2{s} \cdot {\pi}
\end{equation}
where $s$ is the Bloch vector of the density operator $\omega$.

An estimate of the unknown parameters ${s}=[s_1,s_2,s_3]^T$ is obtained
by solving the constraint quadratic optimization problem:
\begin{eqnarray}
          \mathrm{Minimize} &&  \qquad L(\omega)  \label{opt_lf} \\
 \mathrm{subject~to} & &  \qquad \|{s}\|  \leq 1  \label{opt_constr}
\end{eqnarray}
The above loss function is rather simple and we can solve the constrained
minimization problem explicitly. In the unconstrained minimization,
two cases are possible. First,
$\|\pi\| \le 1$, and in this case the constrained minimum is taken at $s=\pi$.
When the unconstrained minimum is at $\pi$ with $\|\pi\| > 1$,
then it is clear from the 3-dimensional geometry that the constrained
minimum is taken at
\begin{equation} \label{LS_reg}
        s = \frac{\pi}{\|{\bf \pi}\|}.
\end{equation}

\section{Simulation experiments}

The aim of the experiments is to compare the properties of the above
described least squares and Bayesian qubit state estimation methods.

The base data of the estimation is obtained by measuring spin
components $\sigma_1, \sigma_2,$ and $\sigma_3$ of several qubits
being in the same state i.e. having
just the same Bloch vector $s$. The number of the measurements of
each direction is denoted by $n$ in what follows. The same measurement
data had been used for the two methods. The Bayesian method was applied
with conditioning and also without it to analyze its effect.

The measurements were performed on a quantum simulator for two level
systems implemented in MATLAB\cite{mathworks}.
An experiment setup
consisted of a Bloch vector $s$ to be estimated and a number of spin
measurements performed on the quantum system. The internal
random number generator of MATLAB was used to generate "measured
values" according to the probability distribution of the measured
outcomes. In this way a realization of the random measured data set is
obtained each time we run the simulator.
Each experiment setup was used five times and the \emph{performance indicator
 quantities}, the fidelity, the Hilbert-Smith norm of the estimation error
  and the empirical variance of the estimate were averaged.

\section{Results of the experiments}

The fidelity (\ref{E:fidelity}) of the real Bloch vector and the estimated one,
variance of the estimations (\ref{E:variance}), and the Hilbert-Schmidt
norm (\ref{E:HS}) of the estimation error were the quantities which have
been used to indicate the performance of the methods.

\subsection{Number of measurements}
The first set of experiments were to investigate the dependence
between the performance indicator quantities and the number of measurements $n$.

\begin{figure}[!ht]
\center{\includegraphics[width=5.5cm,keepaspectratio]{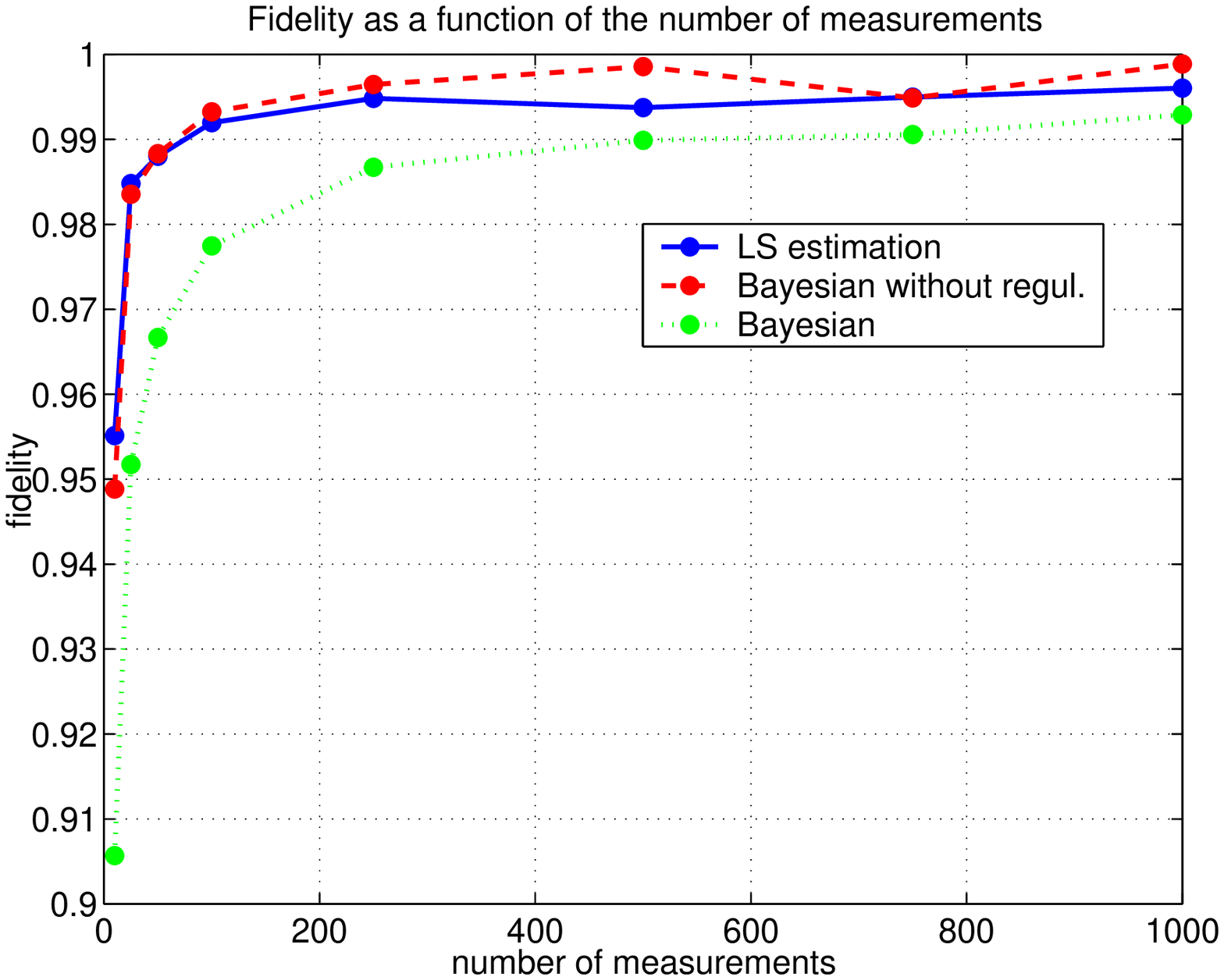}
\includegraphics[width=5.5cm,keepaspectratio]{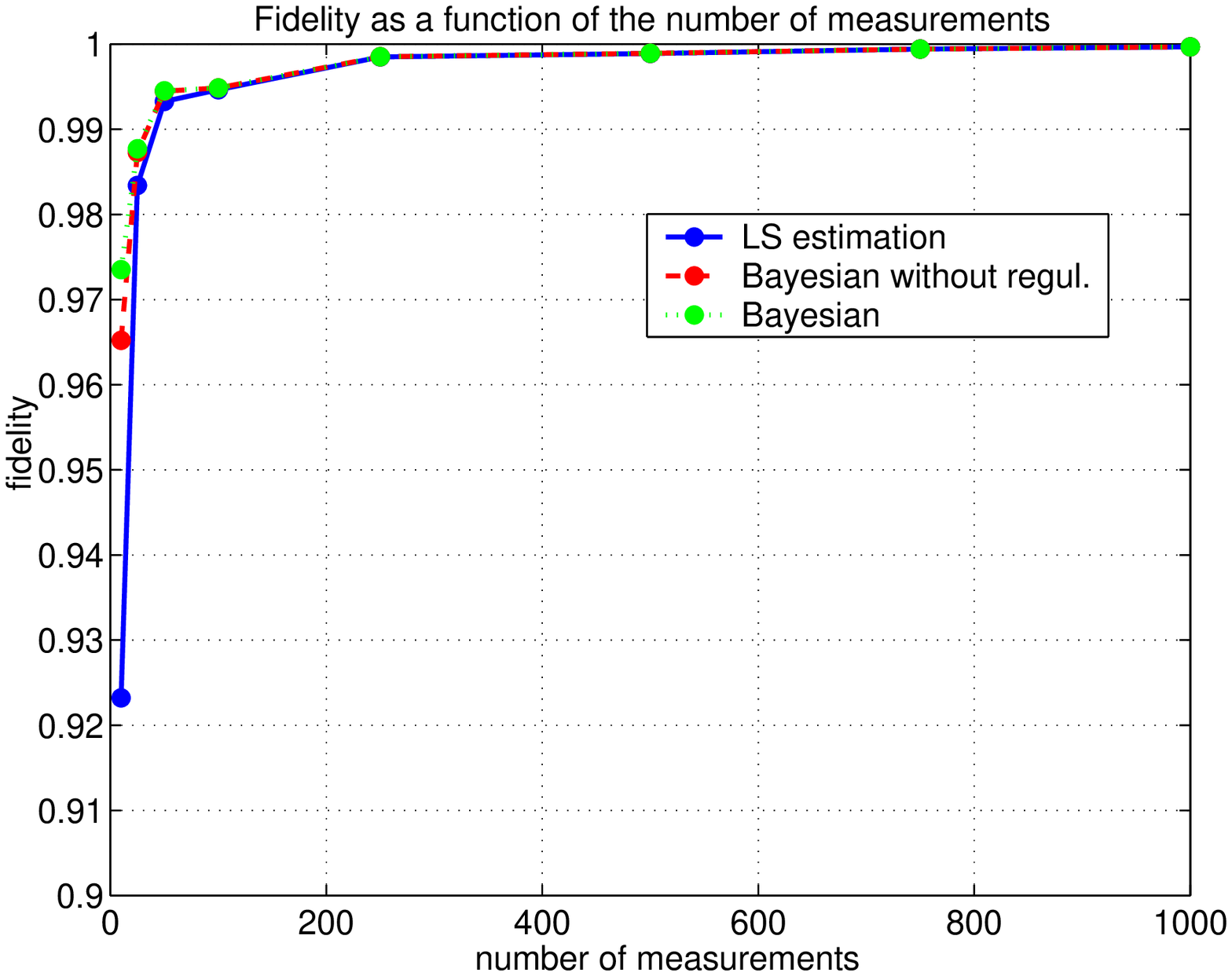}}
 \caption{Fidelity as a function of $n$ for a pure state ($s_{pure}$) and
a mixed state ($s_{mixed}$)}
 \label{fig:fid_n_pure}
\end{figure}

\textbf{Fidelity.} It was expected that
the fidelity goes to 1 when $n$ goes to infinity. Fig.
\ref{fig:fid_n_pure} shows the experimental results for estimating a
pure state ${s_{pure}}=[0.5774, \,0.5774, \, \allowbreak 0.5774]^T$.
\noindent
The result of the Bayesian estimation (dotted line)
shows the weakest performance because of the conditioning feature of
the method: the conditioned joint probability density function
gives worse estimation, than the original one (dashed line). On
the other hand, the original Bayesian without conditioning tends to
give defective Bloch vector estimates with length greater than one.
The price of the validity of the Bayesian method with conditioning
is the precision for (near) pure states.
It is apparent that the least squares
estimation does not have the above problem.

The situation is a little bit different for estimating mixed states
(${s_{mixed}} = [0.3, -0.4, \allowbreak 0.3]^T$). It can be seen that the two
kinds of Bayesian estimation differ only for small $n$'s. When $n$
is greater than $25$, the conditioning has no traceable
effect, i.e. the Bayesian estimation with and without conditioning
gives the same result. Least squares method also works a little bit better for
mixed states than for pure states, at least for larger $n$'s.
It can be seen that pure states
 are a challenge for both methods but least squares handles this difficulty
 a bit better.

\begin{figure}[!ht]
\center{\includegraphics[width=5.5cm,keepaspectratio]{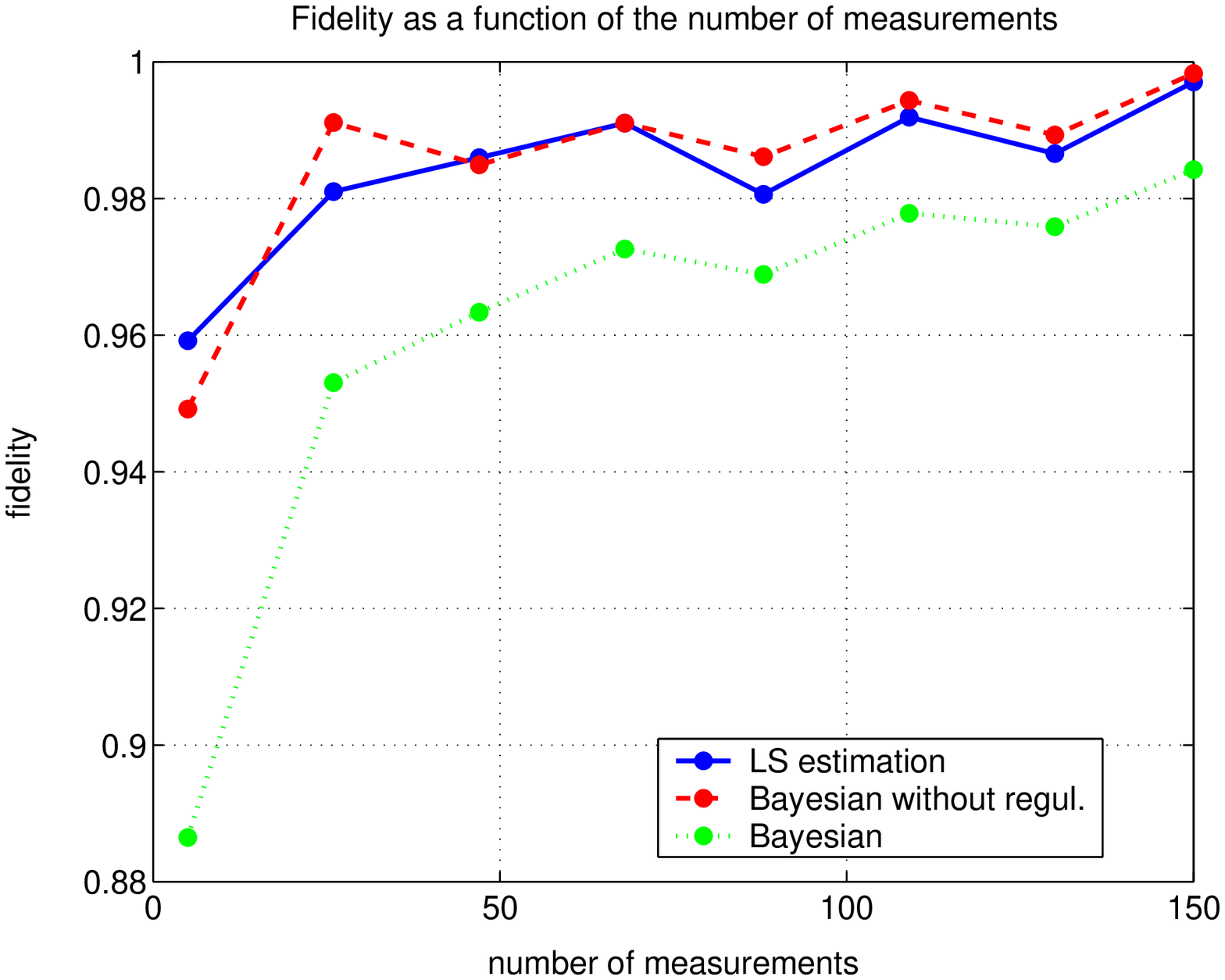}
\includegraphics[width=5.5cm,keepaspectratio]{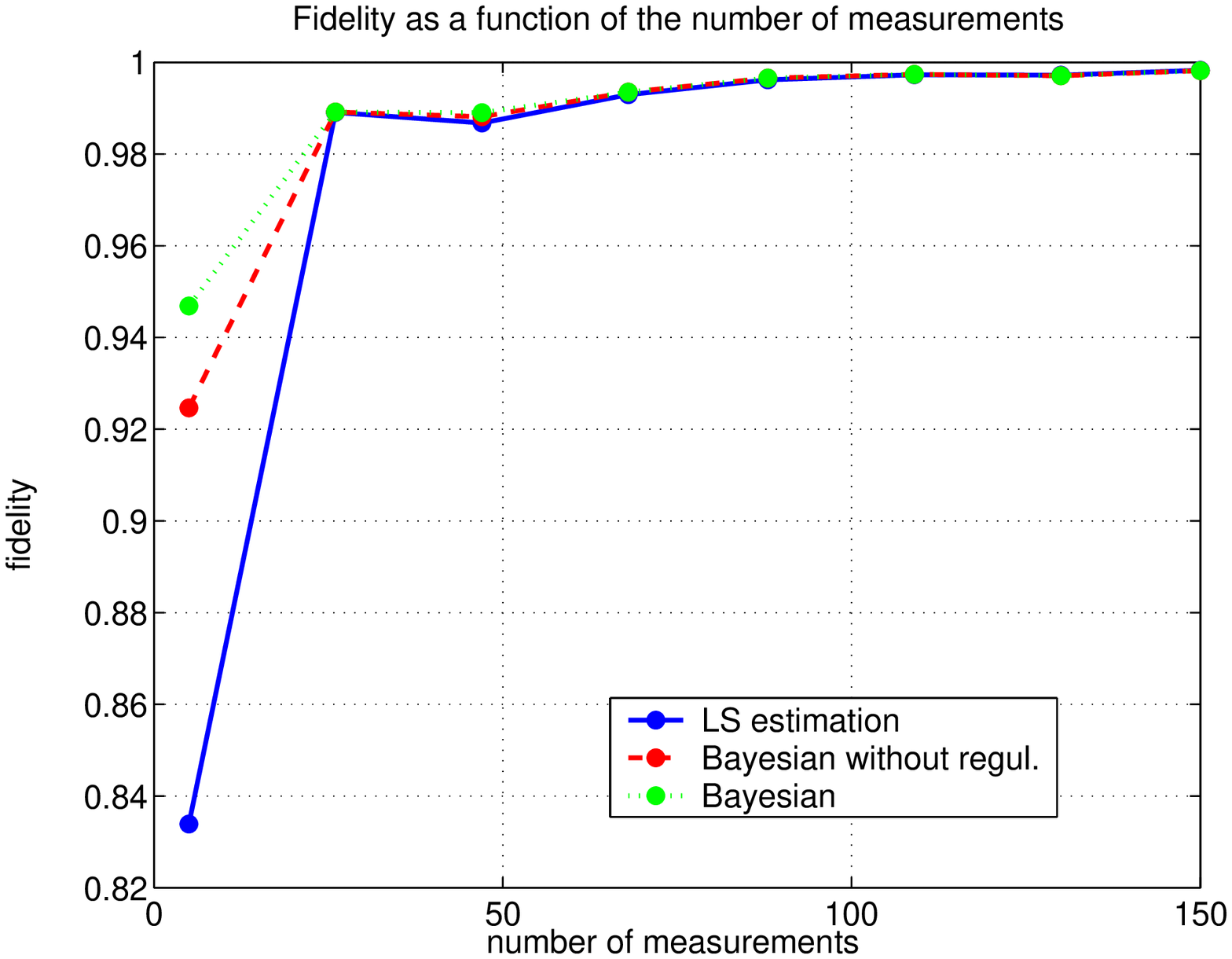}}
 \caption{Fidelity as a function of low $n$ for a pure state ($s_{pure}$) and
a mixed state ($s_{mixed}$)}
 \label{fig:fid_nlow_pure}
\end{figure}

In order to investigate more deeply the behavior of the estimates with low
 number of measurements we show the variation of the fidelities
 as a function of the number of measurements in the interval $n=[5,150]$
 for both the pure and mixed states above (see Fig. \ref{fig:fid_nlow_pure}).
It was expected that the Bayesian estimates outperform the LS one for low
 number of experiments, but it is only true in the case of mixed states.
For pure states the overly conservative conditioning of the Bayes method
 causes a bias.
In addition, one can notice, that the effects related to the low number
 of measurements can be seen only when $n<25$.

\begin{figure}[!ht]
\center{\includegraphics[width=5.5cm,
keepaspectratio]{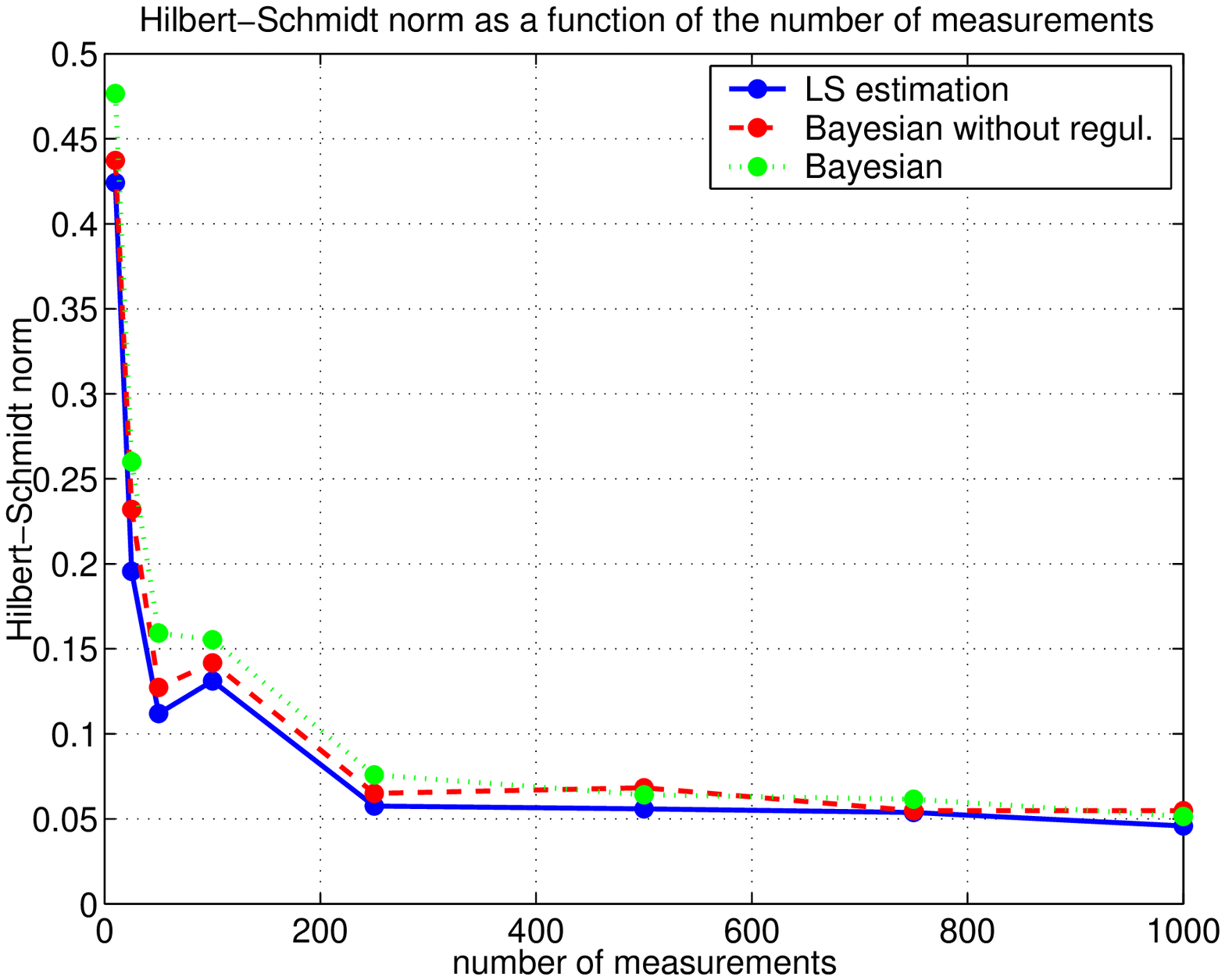}\includegraphics[width=5.5cm,keepaspectratio]{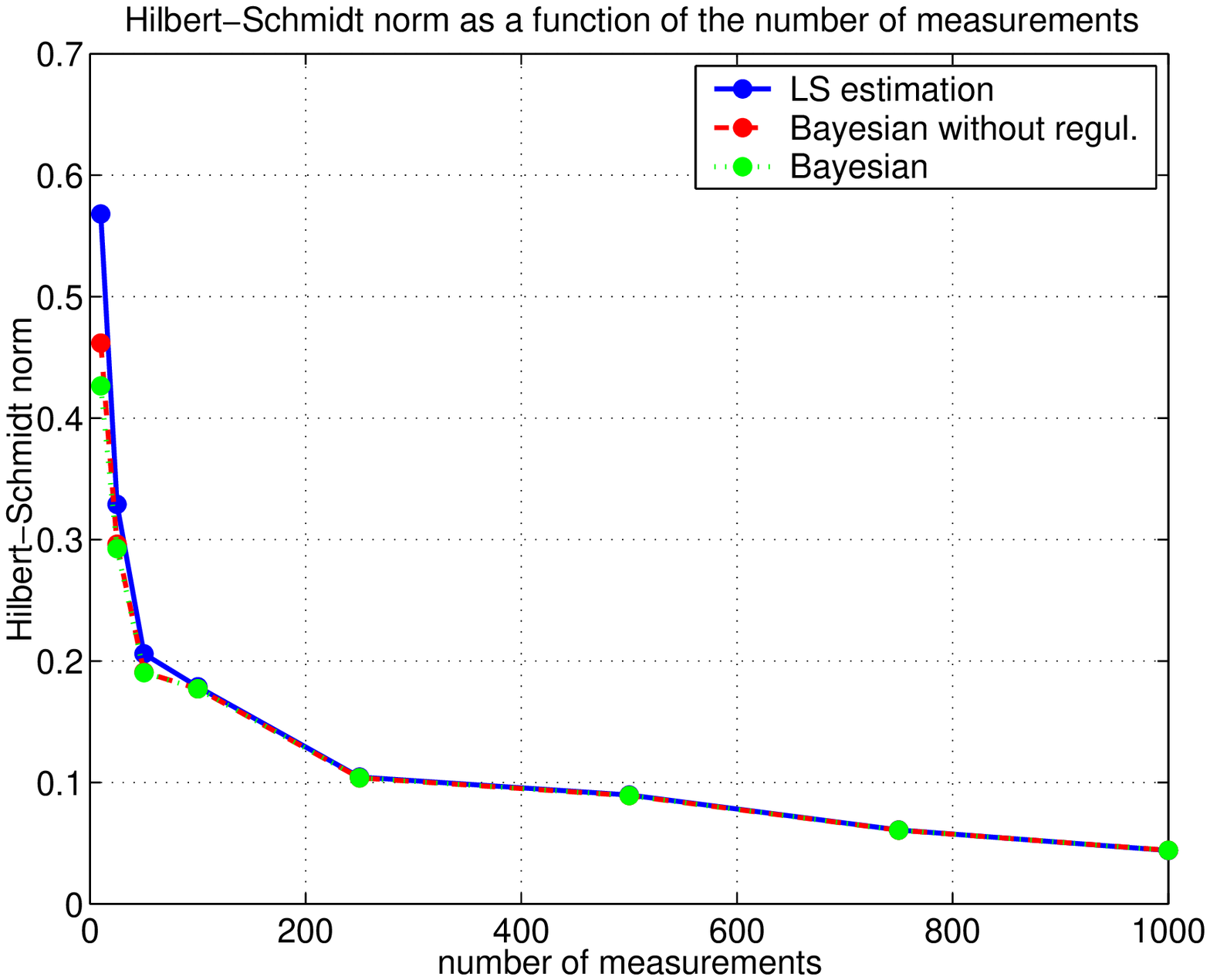}}
 \caption{The Hilbert-Schmidt norm as a function of $n$ for a pure state
($s_{pure}$) and a mixed state ($s_{mixed}$)}
 \label{fig:L2_n_pure}
\end{figure}

\begin{figure}[!ht]
\center{\includegraphics[width=5.5cm,
keepaspectratio]{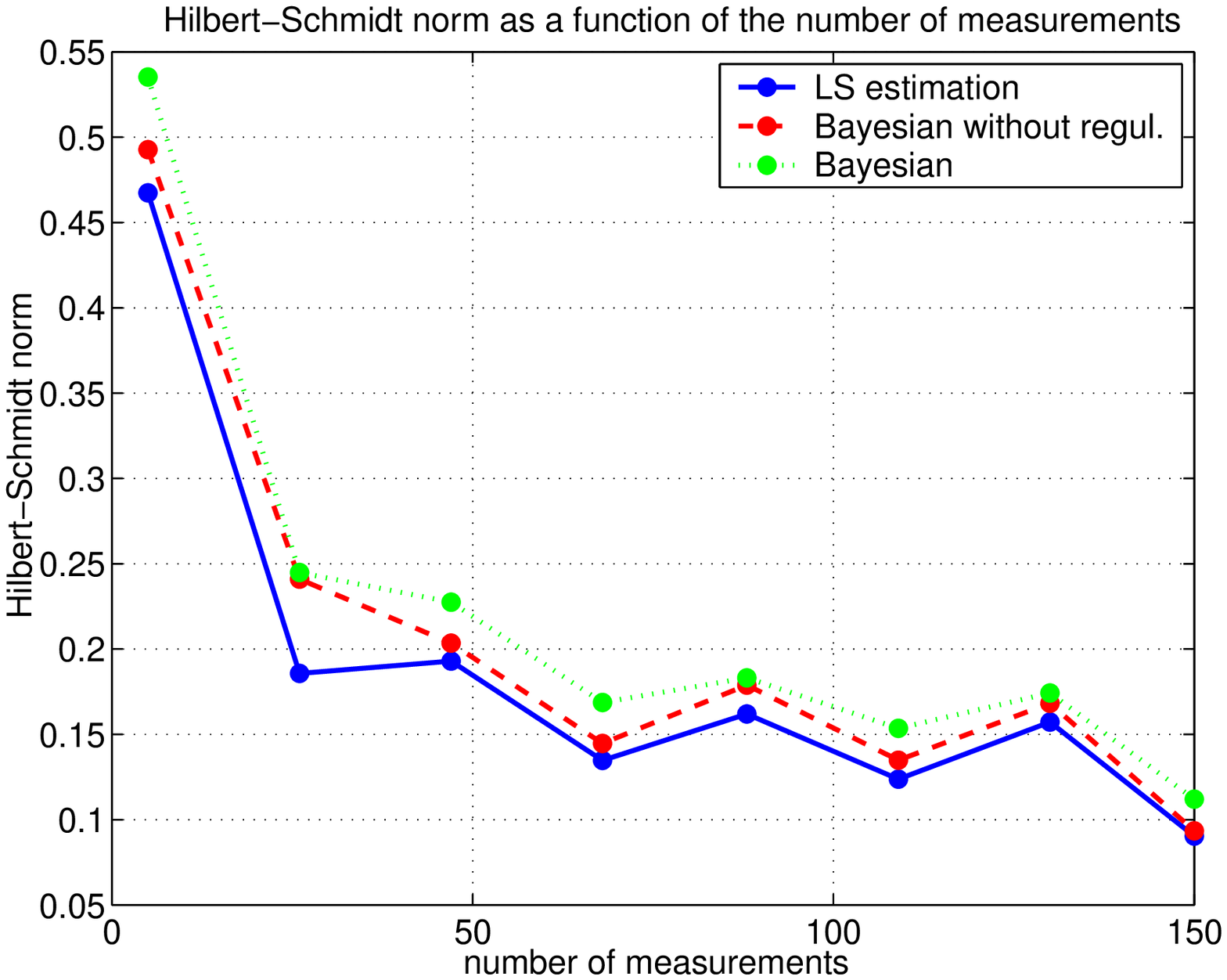}\includegraphics[width=5.5cm,keepaspectratio]{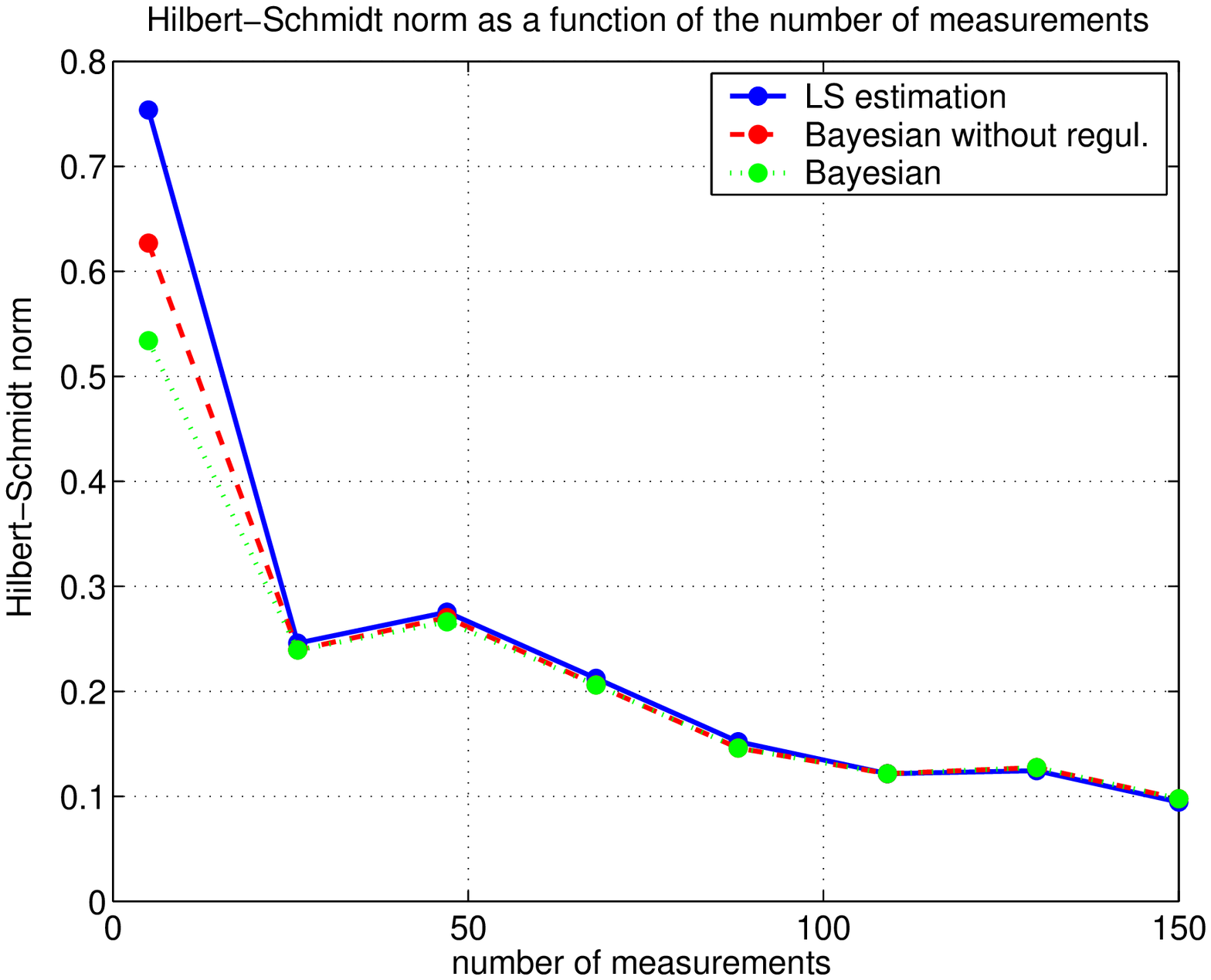}}
 \caption{The Hilbert-Schmidt norm as a function of low $n$ for a pure state
($s_{pure}$) and a mixed state ($s_{mixed}$)}
 \label{fig:L2_nlow_pure}
\end{figure}

\textbf{Hilbert-Schmidt norm.}
For Hilbert-Schmidt norm, it was expected to decrease to
 zero in the limit. The experiments seem to come up to expectations
(Fig. \ref{fig:L2_n_pure}). In the case of pure states the same
phenomena is noticeable as for fidelity.
If one zooms on the low number of measurement region in  Fig. \ref{fig:L2_n_pure}
 then the picture in Fig. \ref{fig:L2_nlow_pure} results.
Here we can see the same effects as for the fidelity, but in a less exposed
 way.
Thus fidelity seems to be a more sensitive indicator of performance than the
 Hilbert-Schmidt norm.

\begin{figure}[!ht]
\center{\includegraphics[width=5.5cm,
keepaspectratio]{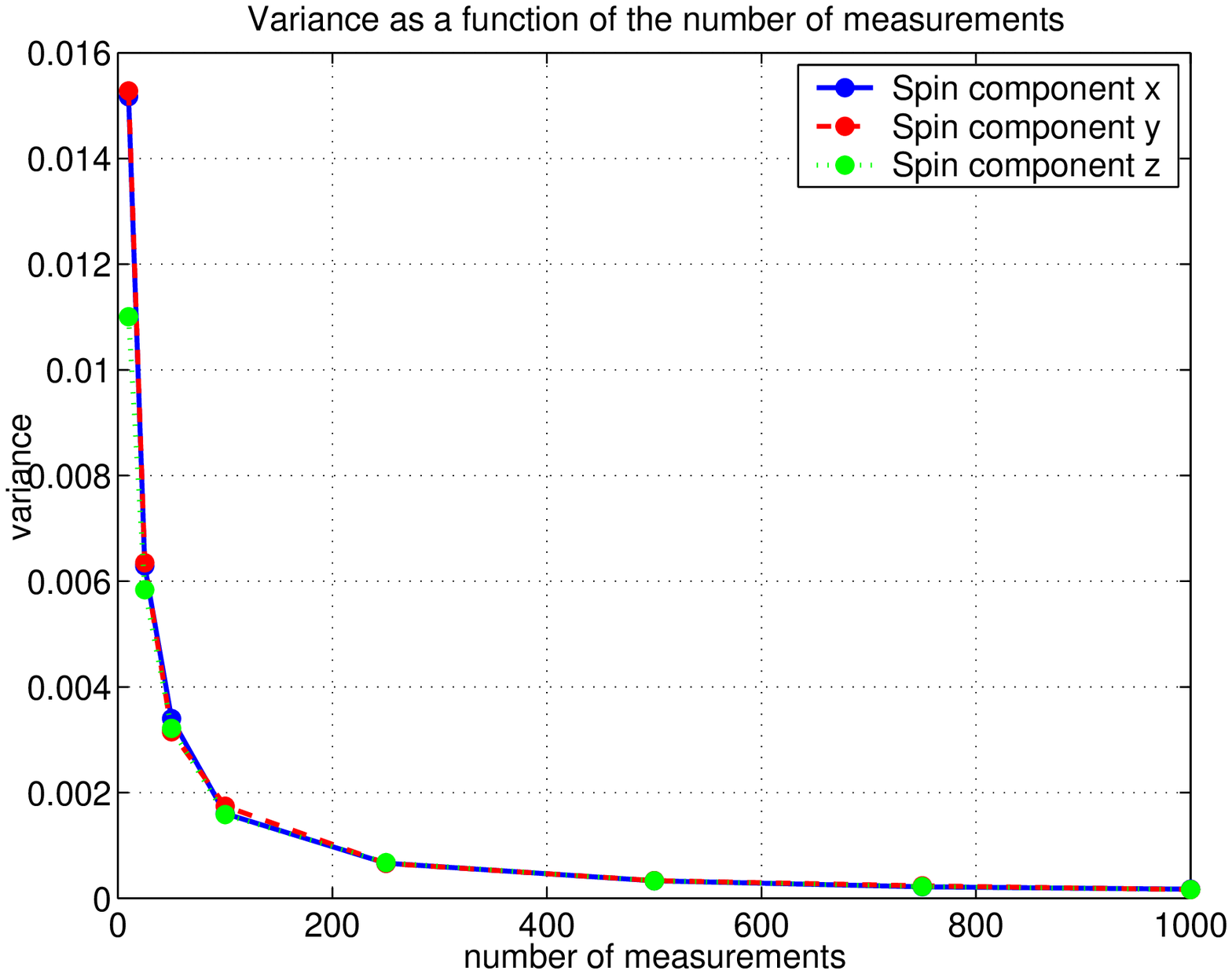}\includegraphics[width=5.5cm,
keepaspectratio]{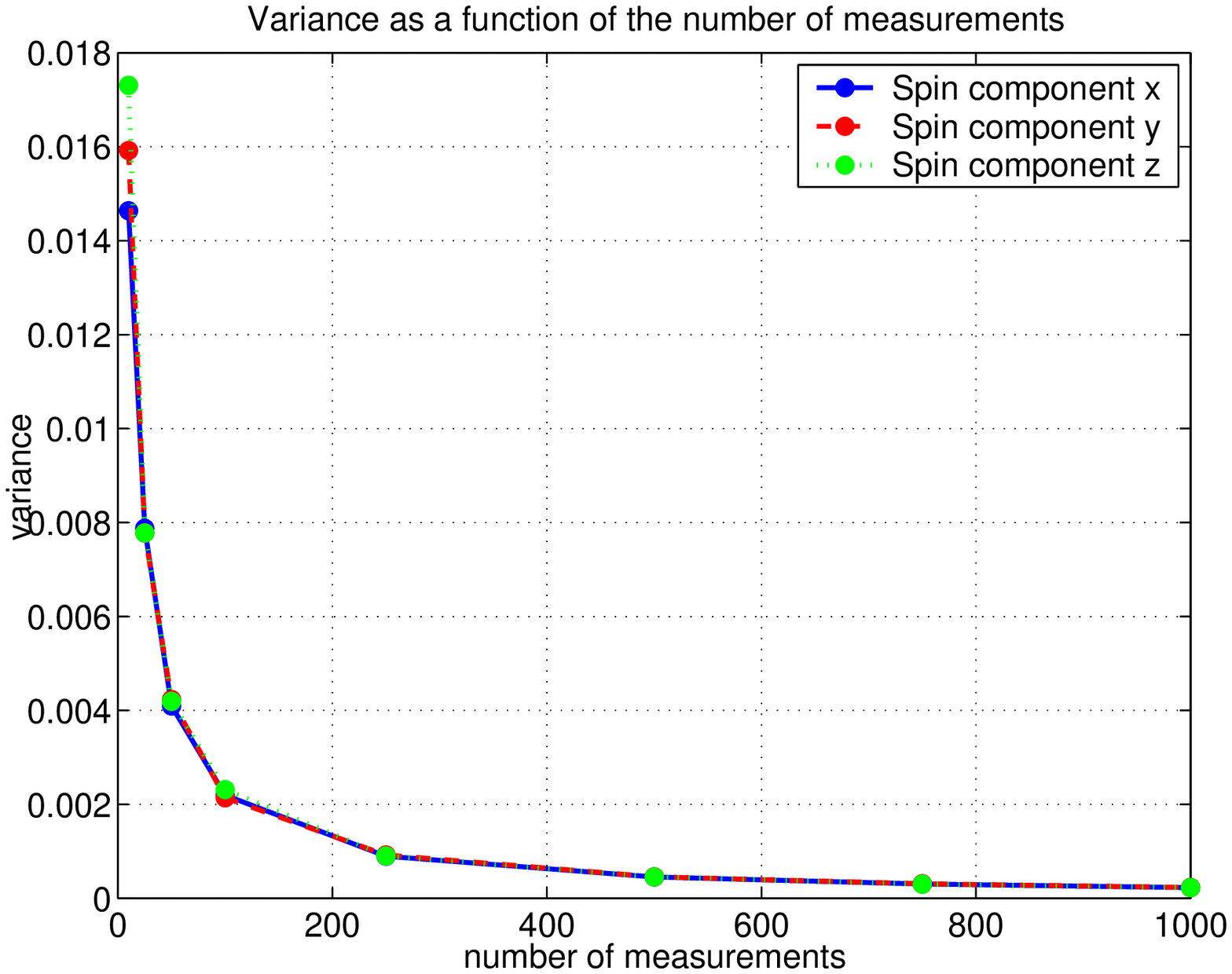}}
\caption{Variance as a function of $n$ for a pure state ($s_{pure}$)
and a mixed state ($s_{mixed}$)}
 \label{fig:var_n}
\end{figure}

\begin{figure}[!ht]
\center{\includegraphics[width=5.5cm,
keepaspectratio]{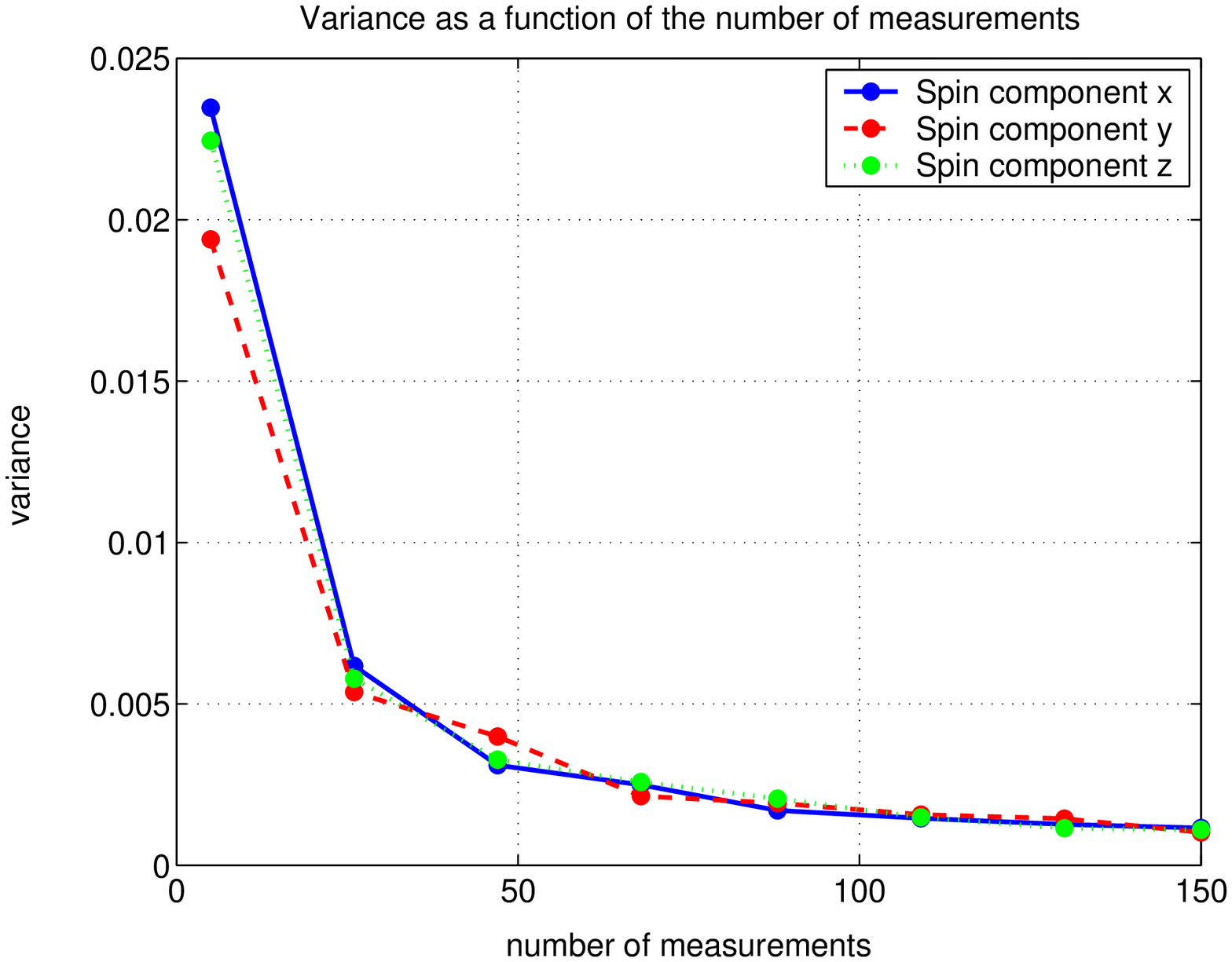}\includegraphics[width=5.5cm,
keepaspectratio]{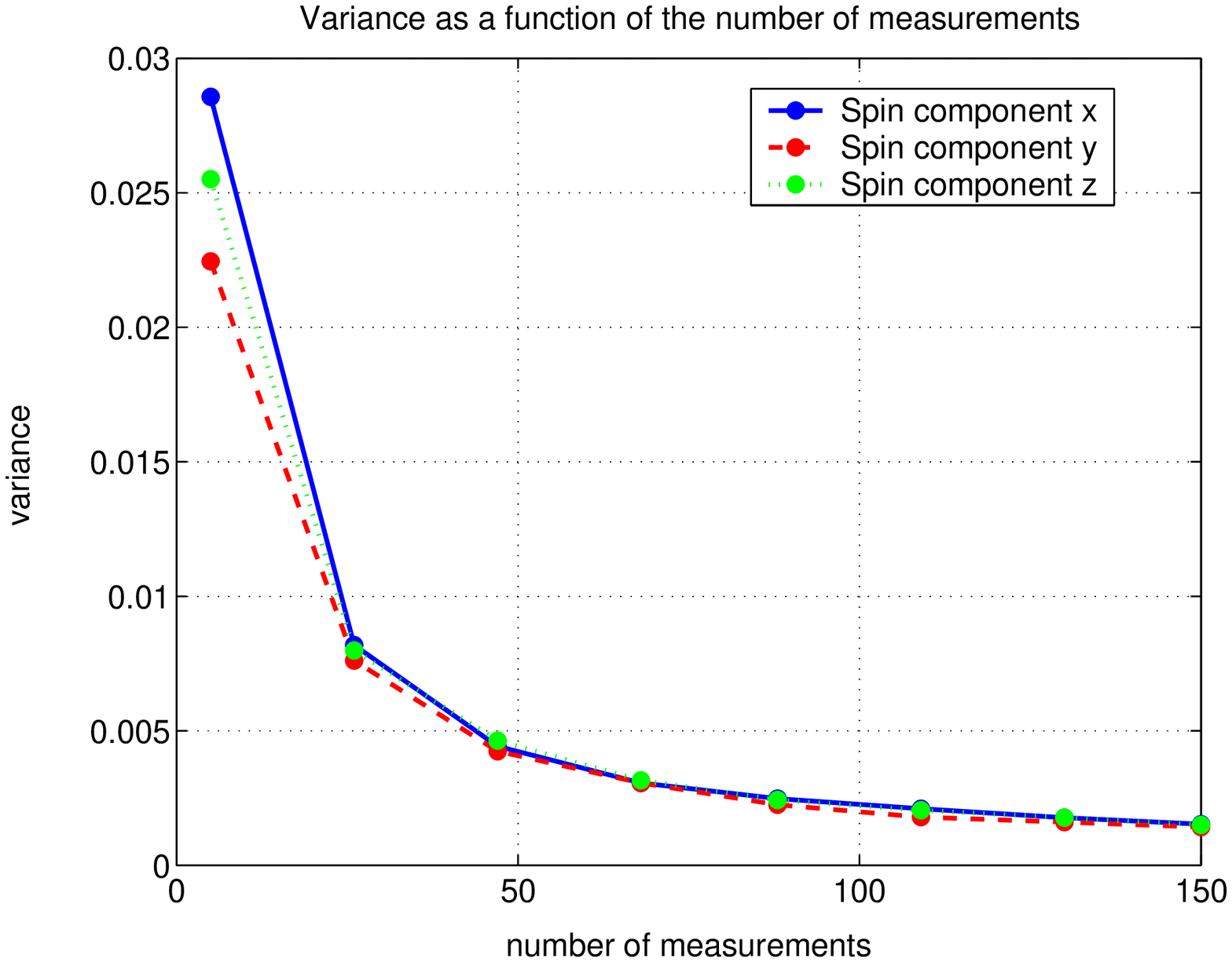}}
\caption{Variance as a function of low $n$ for a pure state ($s_{pure}$)
and a mixed state ($s_{mixed}$)}
 \label{fig:var_nlow}
\end{figure}

\textbf{Variance.}
The variance of the estimates were computed for the Bayesian
estimation before conditioning. As it was expected, there is no
apparent difference between the variance for the three spin
components $s_1, s_2,$ and $s_3$ and the variance decreases with
$n$. The fact that the state to be estimated is a pure or a mixed
state also does not have any effect on the result (Fig. \ref{fig:var_n}).
The same effect can be seen if one focuses on the low number of
 measurement region, as seen in Fig. \ref{fig:var_nlow}.

\subsection{The length of the Bloch vector}

\begin{figure}[!ht]
\center{\includegraphics[width=5.5cm,keepaspectratio]{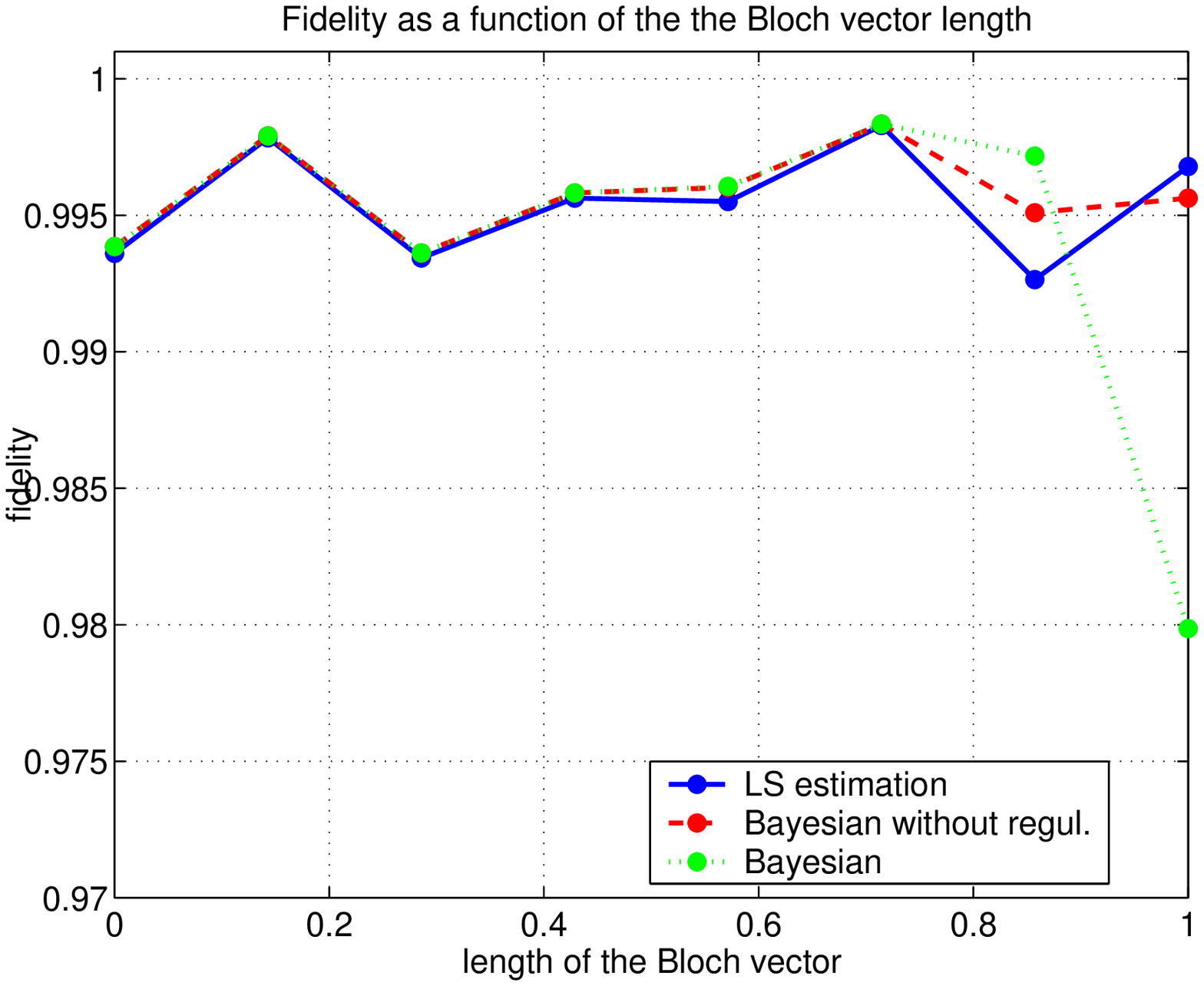}
\includegraphics[width=5.5cm,keepaspectratio]{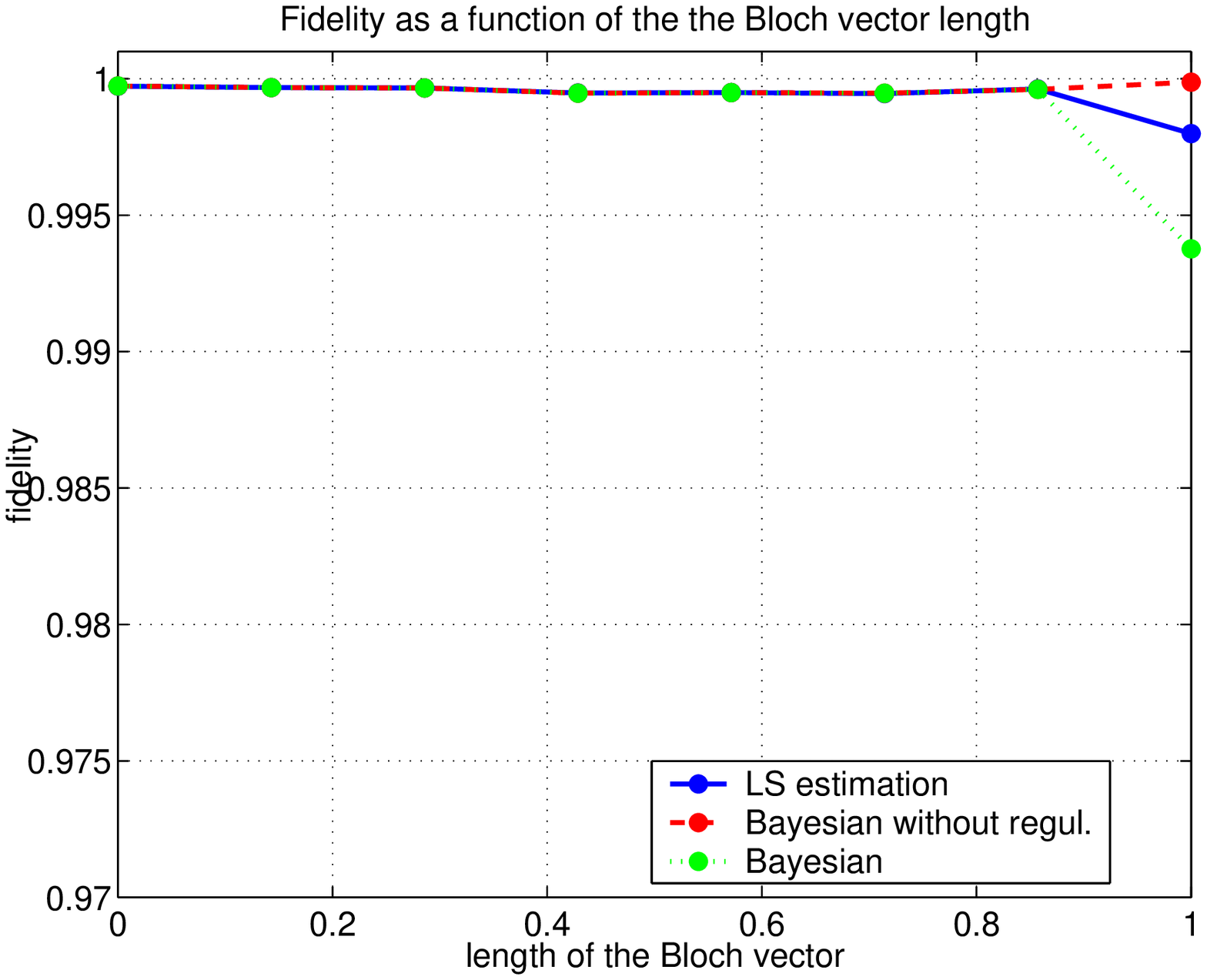}}
 \caption{Fidelity as a function of $\|s\|$ for $n=100$ and $n=900$}
 \label{fig:fid_l}
\end{figure}

\begin{figure}[!ht]
\center{\includegraphics[width=5.5cm,keepaspectratio]{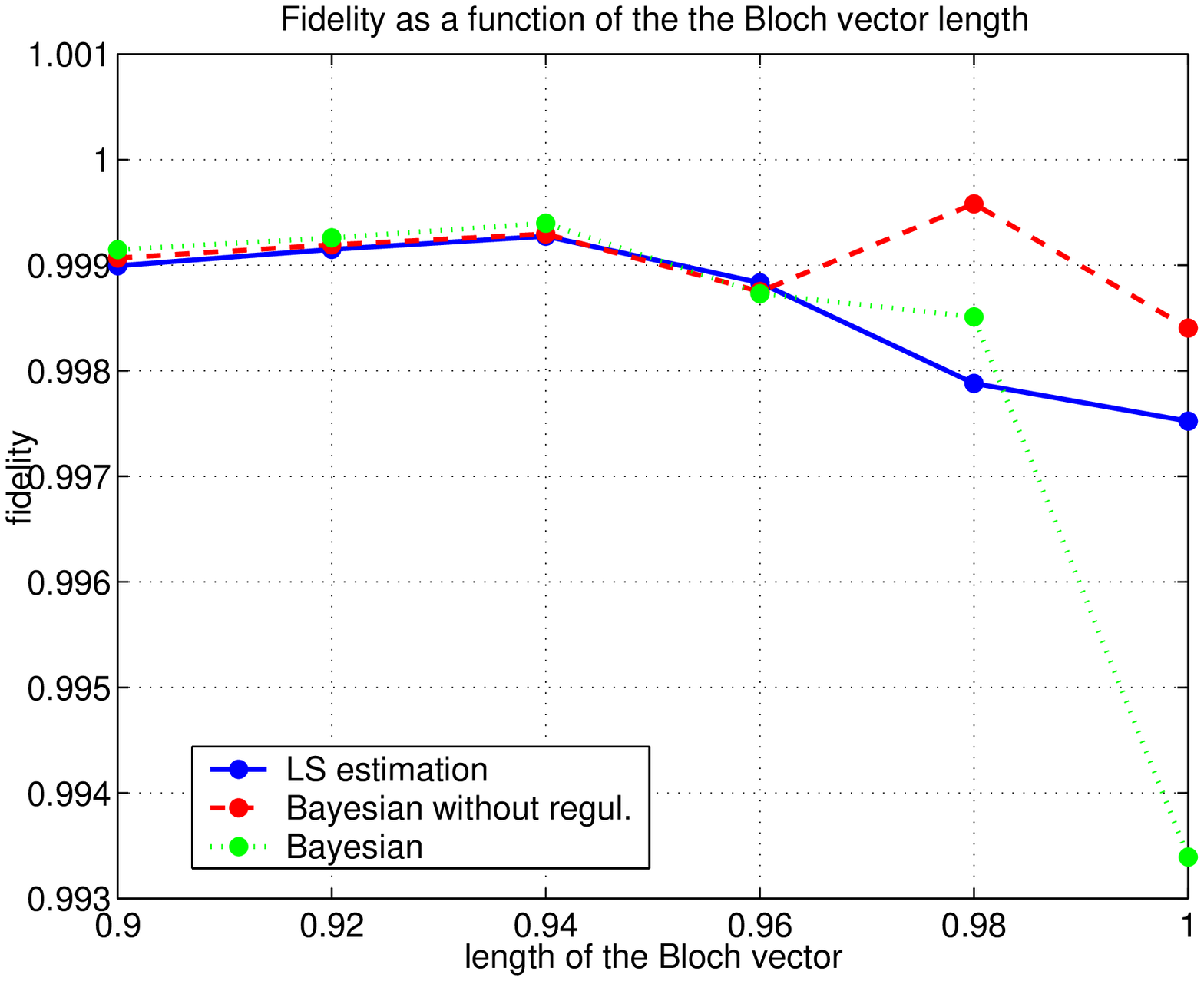}}
 \caption{Fidelity as a function of $\|s\|$ for $n=900$}
 \label{fig:fid_lz}
\end{figure}

During the second set of experiments the length of the Bloch vector
was varying. Its direction was ${s}=[0.5774,\, 0.5774,\, 0.5774]^T$.
The expectation to fidelity was to be relatively independent of the
Bloch vector length $\|s\|$. The experiment results can be seen in
Fig. \ref{fig:fid_l}. The first picture shows the case $n=100$, where, in
spite of the big variance, the conditioned Bayesian shows an increase
near the pure state ($\|s\|=1$). At $n=900$ it is more apparent that LS
and conditioned Bayesian methods (both have certain conditioning
feature to avoid faulty estimates near $\|s\|=1$, see (\ref{LS_reg}),
(\ref{mean_reg})) have worse
performance near pure states.
Fig. \ref{fig:fid_lz} shows fidelity between $\|s\|=0.9$ and $\|s\|=1$ for
$n=900$, where the above mentioned phenomena can be seen more clearly.
\begin{figure}[!ht]
\center{\includegraphics[width=5.5cm,keepaspectratio]{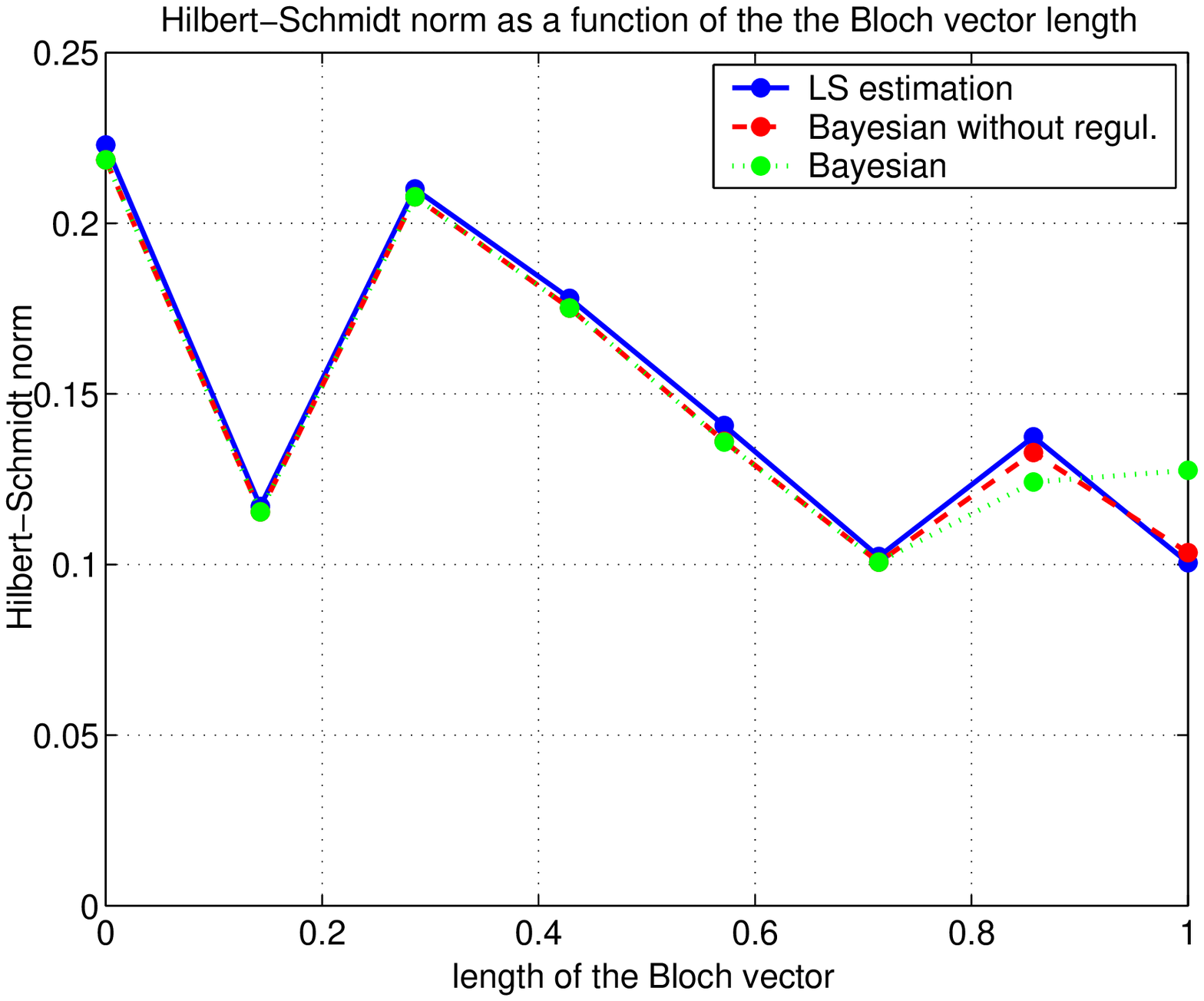}
\includegraphics[width=5.5cm,keepaspectratio]{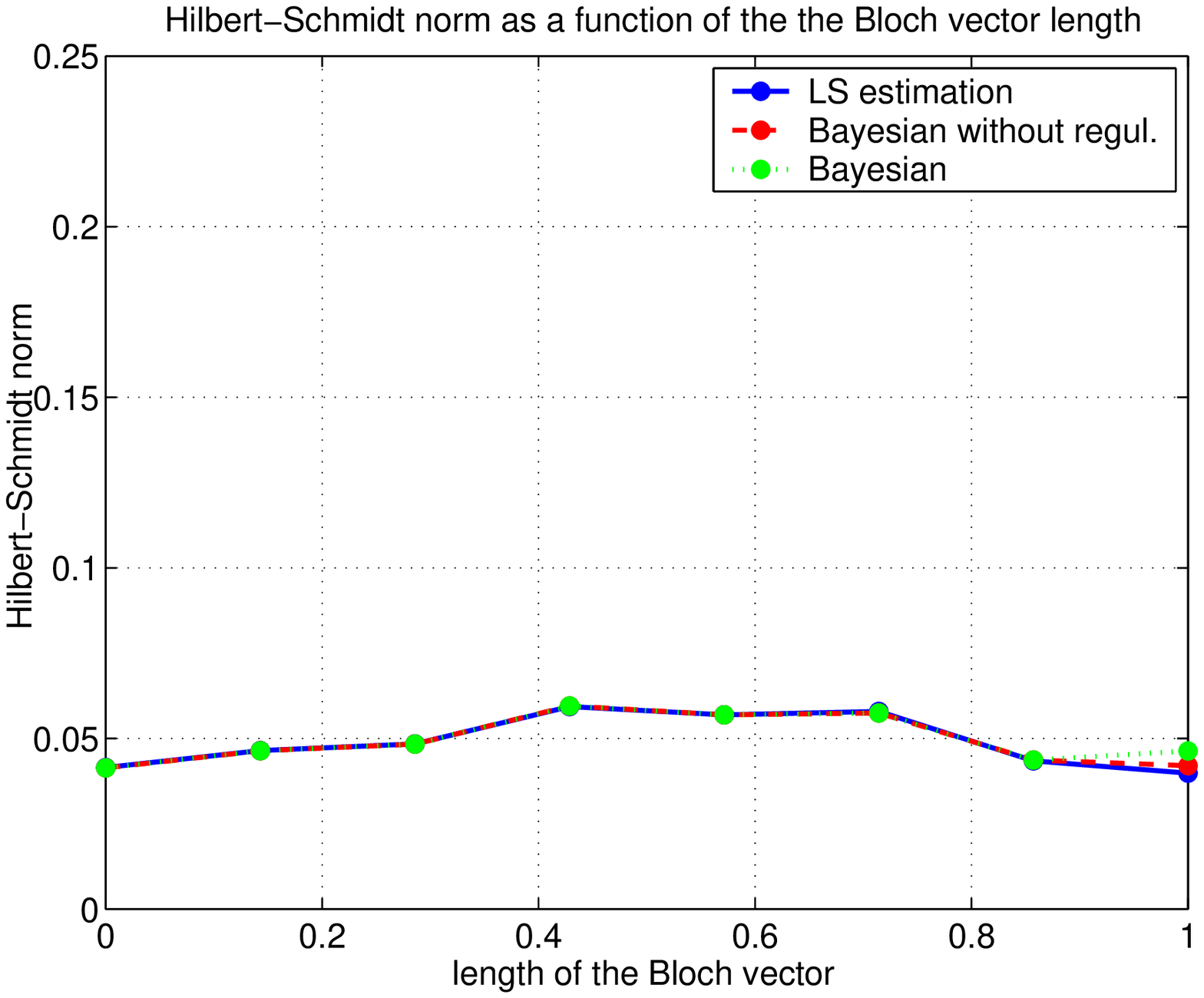}}
 \caption{Hilbert-Schmidt norm  as a function of  $\|s\|$ for $n=100$ and $n=900$}
 \label{fig:L2_l}
\end{figure}

As it was expected, the Hilbert-Schmidt norm seems to be constant for varying
Bloch vector lengths, Fig. \ref{fig:L2_l} shows the simulation
results. For relatively small $n$ the variance is rather big but
increasing the number of measurements it can be seen that
the Hilbert-Schmidt norm is almost constant. Near $\|s\|=1$ there is
a small increasing for the conditioned Bayesian method.

\begin{figure}[!ht]
\center{\includegraphics[width=5.5cm,keepaspectratio]{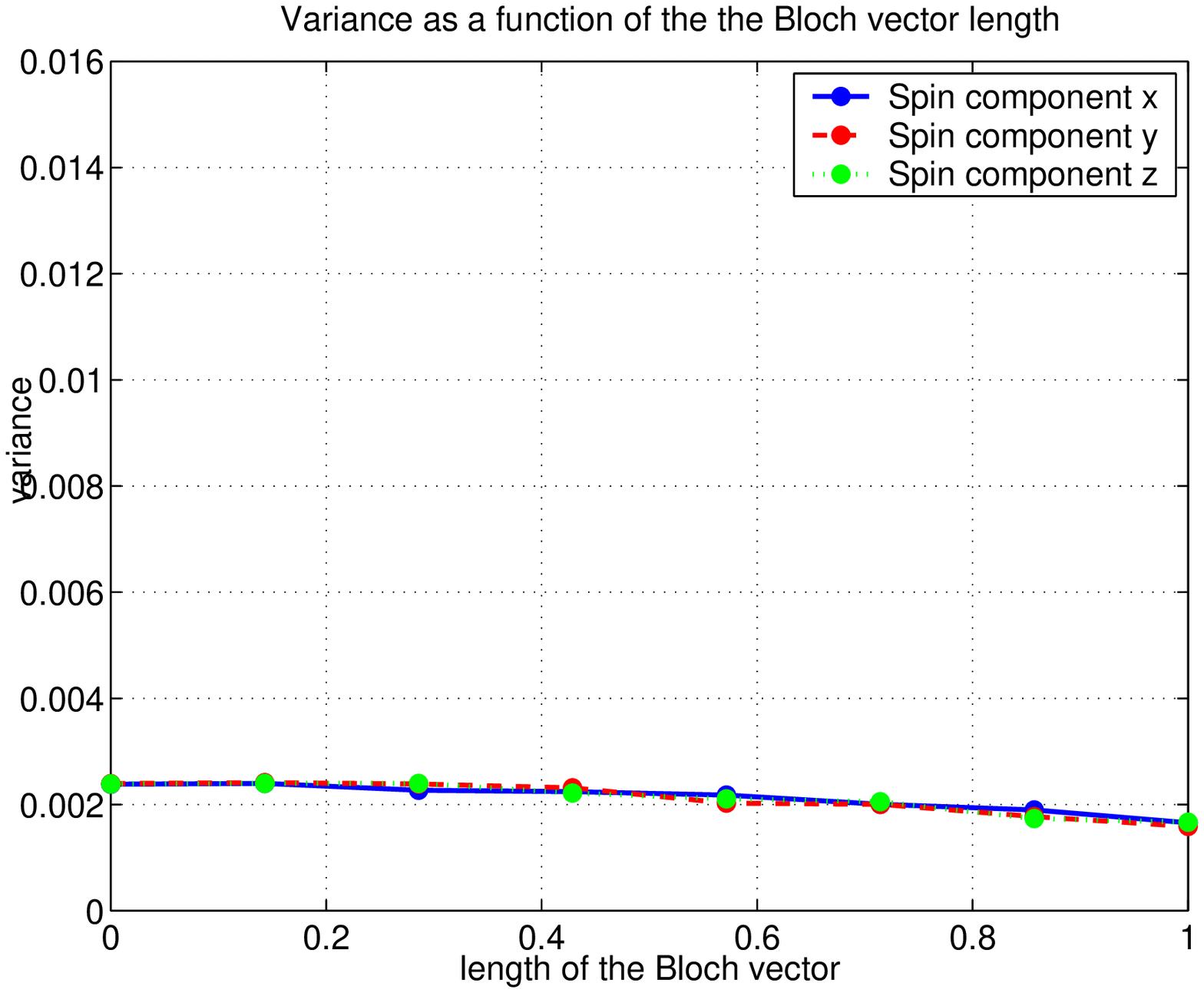}
\includegraphics[width=5.5cm,keepaspectratio]{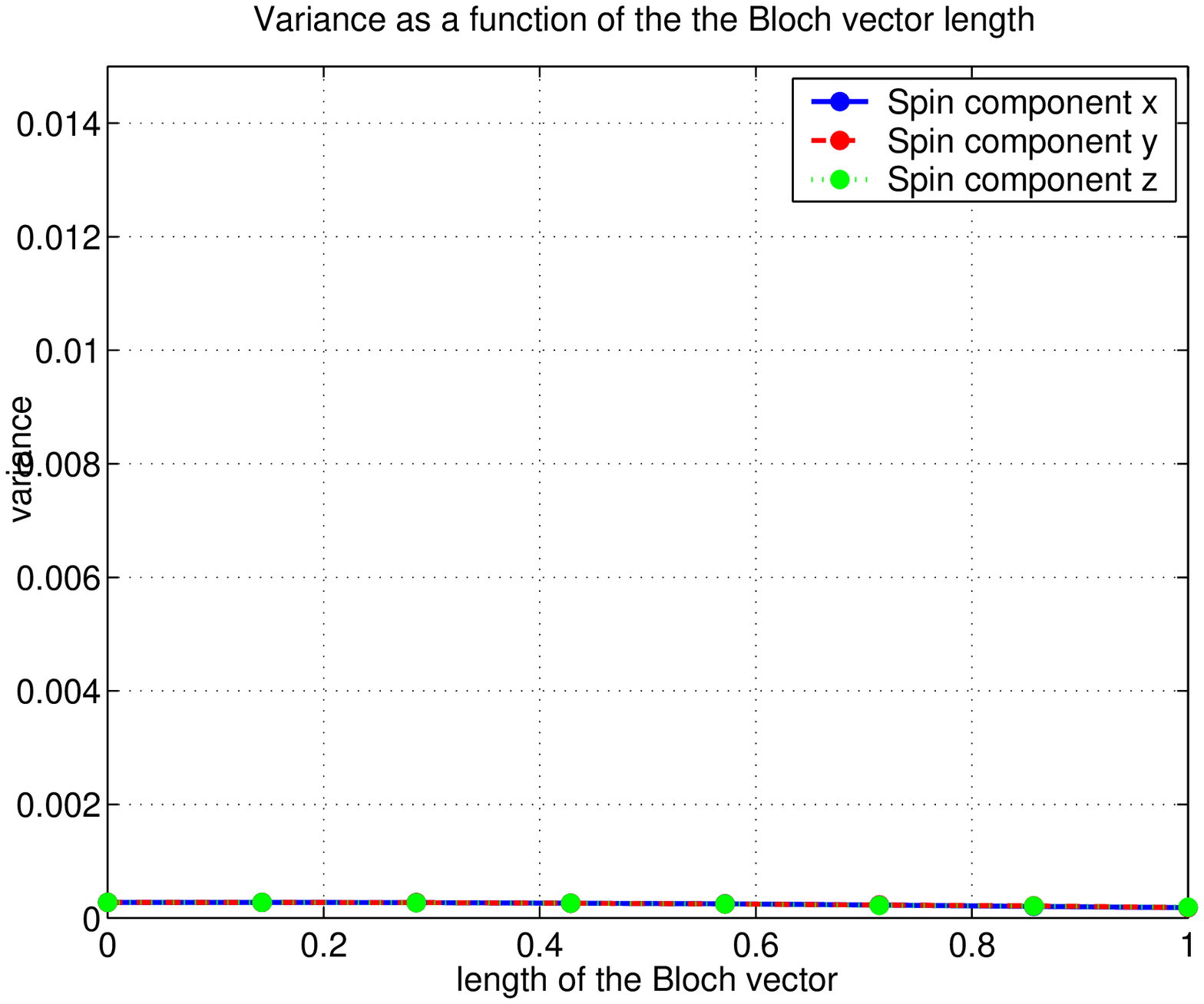}}
 \caption{Variance as a function of $\|s\|$ for $n=100$ and $n=900$}
 \label{fig:var_l}
\end{figure}

The expectation for variance was to be independent of Bloch vector
length. Fig. \ref{fig:var_l} shows the results with the same
variance-scale as in Fig. \ref{fig:var_n}. The first graph is the
results for 100 measurements, the other one is for $n=900$. The
result are in accordance with Fig. \ref{fig:var_n}. As it was
expected, the two graphs can be regarded as constants.

\section{Conclusion}

The performance of two state estimation
 methods, the Bayesian state estimation as a statistical method and the least
 squares (LS) method as an optimization-based method is investigated in this paper
 by using simulation experiments.
The fidelity and the Hilbert-Smith norm of the estimation error
  as well as the empirical variance of the estimate are used as
  performance indicator quantities.
The variation of these quantities as functions of the number of measurements and
 the length of the Bloch vector are computed.

It is found that fidelity is the best indicator for the quality of an estimate
 from the investigated three performance indicator quantities from
 both qualitative and quantitative point of view.
For state estimation of a single qubit the region of the 'low measurement number'
 being $n<25$ and the 'large measurement number' $n>200$ has been determined experimentally.
As for the comparison of the different state estimation methods
 we have found that the Bayesian method could outperform the LS estimation only
 in the case of mixed states for low number of measurements (below $n=25$).

The investigated methods were found to be quite sensitive to the length
 of the Bloch vector, i.e.\ to the fact if a pure or mixed state
 was the one to be estimated.
The methods that are not informed about the purity of the state
 can perform quite badly if they are used to estimate a pure state or a
 "nearly pure" state.

It is also found that the way of conditioning is critical for the methods
 capable of estimating both
 pure and mixed states.
The simple length constraint of the least squares method (in (\ref{LS_reg}))
 seems to work quite effectively, thus a version of the
 Bayesian estimation method with LS-type
 constraining is a good candidate of an improved stochastic
 state estimation method.

To handle somehow the difficulties related to estimating nearly
 pure states one should avoid to use a flat geometry on the state space
  but one should probably use a suitably defined special Riemannian 
geometry instead.

\bibliography{QUANTUM}

\end{document}